\documentstyle[preprint,aps,psfig]{revtex}
\def\bea{\begin{equation}}
\def\eea{\end{equation}}
\begin{document}
\draft
 \preprint{NT@UW-99-25}
\title{  
Perturbative Pion Wave function in Coherent Pion-Nucleon Di-Jet  Production}
\author{L. Frankfurt}
\address{ School of Physics and Astronomy, \\ 
Tel Aviv University, 69978 
Tel Aviv, Israel}
\author{G. A. Miller}
\address{Department of Physics, Box 351560\\
University of Washington \\
Seattle, WA 98195-1560\\U.S.A.}
\author{M. Strikman}
\address{ Department of Physics,
Pennsylvania State University,\\
University Park, PA  16802, USA} 
\maketitle
\begin{abstract}
  A perturbative QCD  treatment 
of the pion wave function is applied to computing the scattering amplitude
  for coherent 
high relative momentum 
di-jet production from a  nucleon. 
  \end{abstract}

\section{Introduction}
Consider  a process in which a high momentum  ($\sim$ 500 GeV/c)  pion
undergoes a coherent interaction 
 with a nucleus in  such a way  that the final state
consists of two jets (JJ) moving at high transverse relative
momentum ($\kappa_\perp>1\sim2 $ GeV/c). In this coherent
process, the final nucleus is in its ground  state. This  process is very
rare, but it
 has remarkable properties\cite{fms93}. The selection of the final state 
to be  a $q\bar q$ pair plus the nuclear ground state 
causes the $q\bar q$ component of 
the pion dominate the reaction process.  At very  high beam momenta, 
the pion breaks up into a $q\bar q$ pair well before hitting the nucleus.
Since the momentum transfer
to the nucleus is very small (almost zero for forward scattering), the only
source of high momentum is the gluonic interactions between the quark and the
anti-quark. Because $\kappa_\perp$ is large, the quark and anti-quark must
be at small separations--the virtual state of the pion is a
point-like-configuration\cite{fmsrev}. But the coherent interactions
of a color neutral  point-like configuration is suppressed 
by the cancellation of  gluonic emission from the
quark and anti-quark\cite{bb,fmsrev}.  
Thus the interaction with the nucleus is very rare, 
and the pion is most likely to interact with only one nucleon.
For this coherent process, the forward scattering amplitude is almost
(since the momentum transfer is not exactly zero)
proportional
to the number of nucleons, $A$ and the cross section varies as $A^2$.
This reaction,   in which there are no  initial or final state interactions,
is an  example  of 
(color singlet)
color transparency \cite{mueller,fmsrev}.
This is the name given to a high momentum transfer process in which
the normal strongly absorbing  interactions 
 are absent, and the nucleus is transparent.
The term 
``suppression  of a color coherent process" could also be used, because it is
the quantum mechanical
destructive
 interference 
of amplitudes caused by the different color charges of a color singlet
that is responsible for the reduced nuclear interaction.

The forward angular distribution is difficult to observe, so 
one integrates the angular distribution, and  the  $A^2$ variation becomes
$\approx A^{4/3}$. 
But  the inclusion of the leading correction
to this process, which arises 
from multiple scatterng of the point-like configuration 
 causes a further increase in the $A$-dependence\cite{fms93}.
Actually at sufficiently small 
$x_N={2\kappa_t^2\over s}\le {1\over 2m_NR_A}$, the 
 situation changes since the quark-antiquark
system scatters off the collective gluon field of the nucleus. Since this 
field is expected to be   shadowed, one expects a gradual  disappearance
of color transparency  for $ x\le 0.01$  - this is the
onset of perturbative color opacity
\cite{fms93}.
Within the kinematical region of applicability of the QCD
 factorization theorem, 
the A dependence of this process is given by the factor:
$A^{4/3}\left[G_A(x,Q^2)/G_N(x,Q^2)\right]^2$ \cite{fms93}

Our interest in this 
 curious process has been renewed recently by experimental
 progress\cite{danny}. The preliminary result from experiments comparing
 Pt and C targets is a dependence $\sim A^{1.55\pm0.05}$, qualitatively similar
 to our 1993  prediction.
It is much stronger than the one observed for the soft diffraction
of pions off nuclei
(for a 
review and references see \cite{fmsrev})
, and it is qualitatively different 
from the behaviour  $\sim A^{1/3}$ suggested in \cite{bb}.

Since 1993  many workers have been able to make
considerable progress in the theory related to
the application of QCD to experimentally relevant observables,
and we wish to incorporate that progress and improve our calculation. 
 Our particular aim here is to use perturbative QCD to compute the relevant
high-$\kappa_t, q\bar q$ component of 
wavefunction of the incident  pion. 
We show here that 
QCD factorization
holds for the leading term which dominates at large enough
 values of $\kappa_\perp$. 

 In the following we discuss the different contributions to the scattering 
amplitudes  as obtained in perturbative QCD.

\section{ Amplitude for $\pi N\to N JJ$ }
Consider  the forward ($t=t_{min}\approx 0$)
amplitude, ${\cal M}$, for coherent di-jet production on a nucleon
$\pi N\to N JJ$\cite{fms93}: 
\bea
{\cal M}(N)=\langle f,\kappa_\perp,x \mid {1\over2}\widehat{f}\mid \pi\rangle,
\label {matel}\eea
where $\widehat{f}$ represents the soft interaction with the target nucleon.
The initial $\mid \pi\rangle$ and final $\mid f, \kappa_\perp x \rangle$
states represent the 
physical states,  which generally involve all manner of multi-quark and gluon
components.
Our notation is that $x$ is the fraction of the total  longitudinal momentum of
the incident pion, and $1-x$ is the fraction carried by the anti-quark.
The transverse momenta are given by $\vec{\kappa}_\perp$
and $-\vec{\kappa}_\perp$.

As discussed in the introduction,
for  large enough values of $\kappa_\perp$,
only the $q\bar q$ components of
the initial pion and final state  wave
functions are relevant in Eq.~(\ref{matel}).
This is because we are considering a coherent nuclear process which leads
to a final state consisting of a quark and anti-quark moving at high relative
transverse momentum. The quark and anti-quark ultimately hadronize at distances
far behind the target, and this part of the process is analyzed 
by the experimentalists using a well-known algorithm\cite{danny}.

We continue by letting the 
  $q\bar q$ part of the Fock space be
represented by $\mid \pi\rangle_{q\bar q}$, then 
\bea
\mid \pi\rangle_{q\bar q} =G_0(\pi) V_{eff}^\pi \mid \pi\rangle_{q\bar q},
\label{pieq}
\eea
where $G_0(\pi)$ is the non-interacting $q\bar q$ Green's function evaluated
at the pion mass:
\bea
\langle p_\perp, y\mid G_0 (\pi)\mid p'_\perp,y'\rangle=
{\delta^{(2)}(p_\perp-p_\perp')\delta (y-y')\over
m_\pi^2- {p_\perp^2 +m_q^2\over y(1-y)}},
\eea
where $m_q$ represents
the quark mass, $y$ and $y'$ represent  the fraction of the
longitudinal momentum 
carried by the quark; and the relative transverse momentum between the quark
and
anti-quark is $p_\perp$
and $ V_{eff}^\pi$ is the complete effective interaction,  which includes the
effects of all Fock-space configurations. A similar equation holds for the
final state:
\bea
\mid f,\kappa_\perp,x\rangle_{q\bar q} =\mid \kappa_\perp,x\rangle 
+G_0(f) V_{eff}^f \mid f,\kappa_\perp,x
\rangle_{q\bar q},\label{fstate}
\eea
\bea
\langle p_\perp, y\mid G_0 (f)\mid p'_\perp,y'\rangle=
{\delta^{(2)}(p_\perp-p_\perp')\delta (y-y')\over
m_f^2- {p_\perp^2 +m_q^2\over y(1-y)}}, \label{gf}
\eea
\bea
m_f^2\equiv {\kappa_\perp^2+m_q^2\over x(1-x)},
  \eea  
in which the first term on the right-hand-side of (\ref{fstate})
is the plane-wave part of the
wave function.

The use of the wave functions (\ref{pieq}) and (\ref{fstate}) in the equation
(\ref{matel}) for the scattering amplitude yields
\begin{eqnarray}
{\cal M}(N)&=&{1\over 2}(T_1+T_2),\nonumber\\
T_1&\equiv& \langle \kappa_\perp,x \mid \widehat{f}\mid \pi\rangle,\quad
T_2\equiv _{q\bar q}\langle f,\kappa_\perp,x \mid
V_{eff}^f G_0(f)\widehat{f} \mid \pi\rangle_{q\bar q}.\label{tdef}
\end{eqnarray}
The term $T_2$ includes the effect of the final state $q\bar q$ interaction;
this was not included in our 1993 calculation\cite{fms93}, but its
importance was stressed in \cite{jm} . We shall first evaluate
$T_1$, and then turn to $T_2$.
\section{Evaluation of $T_1$}
The wave function $\mid \pi\rangle_{q\bar q}$  is
dominated by components in which the separation between the constituents is
of the order of the diameter of the physical pion, but there is a perturbative
tail which accounts for short distance part of the pion wave function. This
perturbative tail is relevant here because we need to take the overlap with
the final state which is constructed from constituents moving at high relative
momentum. If we concentrate on those aspects it is reasonable to consider only
the one gluon exchange contribution $V^g$
to $V_{eff}^\pi$ 
and pursue the Brodsky-Lepage 
analysis \cite{BL} for the evaluation of this
particular component. 
Their use of the  light cone gauge $A^+=0$, simplifies the calculation. We 
also use their normalization and phase-space conventions.

We want to draw attention  to the issue of gauge invariance. The
pion wave function  is not gauge
invariant, but 
 the sum of two-gluon exchange  diagrams for the pion transition to $q\bar q$ 
is gauge invariant. This is because only the imaginary part of the scattering
amplitude  
survives  
 in the sum of diagrams, and  because two exchanged 
gluons are vector particles in a  color singlet state   
as a consequence of  Bose statistics. 
So, in this case,  conservation of color current 
 has the same form as conservation of electric current in QED. We also note,
that  
in the calculation of 
hard high-momentum transfer 
processes, the  $q\bar q$ pair in the non-perturbative pion wave function 
should be considered on 
energy shell. Corrections to this enter as an additional factor of 
$1\over \kappa_t^2$ in the amplitude.

We define the 
non-perturbative part of 
the momentum space wave function
as
\bea
\psi
( l_\perp,y)\equiv \langle l_\perp, y\mid \pi\rangle_{q\bar q}.
\eea
We  use the one-gluon exchange approximation to the exact wave function 
of Eq.~(\ref{pieq}) to obtain an
 approximate wave funtion, $\chi$, valid for large values of 
$k_\perp$.
\begin{eqnarray}
\chi(k_\perp,x)={-4\pi C_F} {1\over m_\pi^2-{ k_\perp^2 +m_q^2\over x(1-x)}}
\int_0^1 dy\int {d^2 l_\perp\over(2\pi)^3} V^g(k_\perp,x;l_\perp,y)
\psi(l_\perp,y)
\end{eqnarray}
with
\begin{eqnarray}
 V^g(k_\perp,x;l_\perp,y)=-4\pi C_F\alpha_s
{\bar u (x,k_\perp)\over\sqrt{x}}\gamma_\mu {u(y,l_\perp)\over \sqrt{y}}
{\bar v (x,-k_\perp)\over\sqrt{1-x}}\gamma_\nu 
{v(1-y,-l_\perp)\over  \sqrt{1-y}}d^{\mu\nu}
\nonumber\\
\times
\left[ {\theta(y-x)\over y-x}
 {1\over m_\pi^2-{k_\perp^2-m_q^2\over x} -{l_\perp^2+m_q^2\over 1-y}}
\right], 
      \end{eqnarray}
and  $C_F={n_c^2-1\over 2n_c}={4\over 3}$. 
The range of integration over  $l_\perp$ is restricted by the
non-perturbative
pion wave function $\psi$. Then we set $l_\perp$ to 0 everywhere in the
spinors and energy denominators and evaluate the strong coupling constant
at $k_\perp^2$:
\bea
\alpha_s(k_\perp^2)={4\pi\over \beta \ln{k_\perp^2\over \Lambda^2}}
,\eea
where $\beta=11-{2\over 3}n_f$. Then
\begin{eqnarray}
V^g(k_\perp,x;l_\perp,y)\approx
{-4\pi C_F\alpha_s(k_\perp^2)\over x(1-x) y(1-y)}V^{BL}(x,y)
\end{eqnarray}  
      where
 $V^{BL}(x,y)$ is the Brodsky-Lepage kernal:
      \bea
V^{BL}(x,y)=2\left[\theta(y-x)x(1-y)+{\Delta\over x-y} +(x\to 1-x, y\to 1-y)
\right],\eea 
with the operator $\Delta$ defined by ${\Delta \over
  x-y}\phi(x)={\phi(x)-\phi(y)\over x-y}$. 
This  kernal includes the effects of vertex and quark mass renormalization. 
The net result for the high $k_\perp$
component of the pion wave function is then
\bea
\chi(k_\perp)={4\pi C_F \alpha_s(k)\over k_\perp^2}
\int_0^1 dy V^{BL}(x,y){\phi(y,k_\perp^2)\over y(1-y)}
\label{pieq1},\eea 
where 
\begin{equation}
\phi(y,q_\perp^2)\equiv
\int {d^2 l_\perp\over(2\pi)^3} \theta(q_\perp^2-l_\perp^2)\psi(l_\perp,y).
\end{equation}


The 
quark distribution amplitude $\phi$ can be obtained using QCD
evolution\cite{BL}.
Furthermore, the analysis of experimental data for virtual Compton scattering
and the pion form factor  performed in \cite{tolya,kroll}
shows that this amplitude  
is not far from 
the asymptotic one for 
$k^2_\perp\ge 2-3$  GeV$^2$
\bea
\phi(x)=a_0x(1-x),
\label{phi}\eea
where $a_0=\sqrt{3}f_\pi$ with $f_\pi\approx 93$ MeV.

Equation  (\ref{pieq1}) represents the high relative momentum  part of the
pion wave function.
Using the asymptotic function (\ref{phi}) in Eq.~(\ref{pieq1})
leads to an expression for 
$\psi(k_\perp,x)\propto x(1-x)/k^2_\perp$
which is of the factorized form used in Ref.~\cite{fms93}.

To  compute the amplitude $T_1$, it is necessary to  specify the
scattering operator $\widehat f.$
For high energy scattering the operator
$\widehat {f}$ changes only the transverse
momentum and therefore 
 in the coordinate space representation $\widehat f$ depends on $b^2$. The
 transverse distance operator $\vec b =
(\vec{b}_{q}-\vec{b}_{\bar{q}})$ is canonically conjugate to
 $\vec{\kappa}_\perp$.
At sufficiently small values of $b$,  
the leading twist effect
and  the dominant term at large $s$ arises from the diagrams when 
pion fragments into two  jets as a result of interactions with
 the two-gluon
component of gluon field of a target, see Figure 1.

The perturbative QCD determination of this interaction
involves a diagram similar to the gluon
fusion contribution to the nucleon sea-quark content
observed
in deep inelastic scattering.
One  calculates the box
diagram for large values of $\kappa_\perp$ using the wave function
of the pion
instead of the vertex for $\gamma^*\to q\bar q$. 
The application of QCD factorization theorem leads 
\cite{fmsrev,bbfs,frs} to
\begin{equation}
  \widehat f(b^2)=i s \frac{\pi^2}{3} b^2 \left[ x_N
G_N(x_N, \lambda/b^2) \right] \alpha_{s}(\lambda/b^2),
\label{eq:1.27}
 \end{equation}
 in which $x_N=2\kappa_\perp^2/s$
where $G_N$ is the gluon distribution function of the nucleon, 
and $\lambda(x=10^{-3})=9$ according to 
Frankfurt, Koepf, and Strikman, \cite{fks}.
Accounting 
 for the difference between the  pion mass and mass of two jet system
requires us  to replace the target gluon distribution  by the
the skewed gluon distribution. The difference 
between both distributions is calculable in QCD using the  QCD
evolution equation for the skewed parton distributions\cite{ffs,fg}.

The most important effect shown in Eq.~(\ref{eq:1.27}) is the $b^2$ dependence
  which shows the diminishing strength of the interaction for small values of
  $b$. 
In the leading order approximation 
 it is legitimate to rewrite $\sigma$ in the form:
\bea
 \widehat {f}(b^2)=is \sigma_0 {b^2\over \langle b_0^2 \rangle} \label{fb2}
 \eea
 in which the logarithmic dependence on $b^2$ is neglected. Our notation is
 that $ \langle b_0^2 \rangle$ represents the pionic average of the
 square of the transverse separation, and
 \bea{\sigma_0 \over \langle b_0^2 \rangle}
 \approx{\pi^2\over 3}
 \alpha_s(\kappa_\perp^2)
[x_NG_N^{(skewed)} (x_N,\kappa_\perp^2)]
. 
\eea
The use of Eq.~(\ref{fb2}) allows a simple evaluation 
of the scattering
amplitude $T_1$ because
 the $b^2$ operator acts as  $-\nabla_{\kappa_\perp}^2$. Using
 Eqs.~(\ref{pieq}) and (\ref{fb2})
 in Eq.~(\ref{tdef}), leads to the result: 
 \bea
 T_1=-4i{\sigma_0\over \langle b^2 \rangle}{4\pi C_F\alpha_s(\kappa_\perp^2)
   \over \kappa_\perp^4}
 \left(1+{1\over \ln{\kappa_\perp^2\over \Lambda^2}}\right)a_0x(1-x).
 \label{t1}
\eea
This is, except for the small $1\over \ln{\kappa_\perp^2\over \Lambda^2}$
correction ($\kappa_\perp\approx 2$ GeV and $\Lambda\approx 0.2$)
arising from taking $-\nabla_{\kappa_\perp}^2$ on $\alpha_s$,
is of the same form as the 
corresponding result of our 1993 paper.
The amplitude of Ref.~\cite{bb} varies as a Gaussian in $\kappa_\perp$.

The $\kappa_\perp$  dependence:
${d\sigma(\kappa_\perp)\over d\kappa_\perp^2}\propto 
{1\over \kappa_\perp^8}$ follows 
from  simple reasoning. The probability to find 
a pion at $b\le {1\over \kappa_\perp}$
 is $ \propto b^2$, while the square of the
total cross section for  small dipole-nucleon interactions
   is $\propto b^4$. Hence the 
cross section of productions of jets with sufficiently large values of
$\kappa_\perp $ is $\propto 
{1\over \kappa_\perp^6}$ leading to a differential cross section 
$\propto {1\over \kappa_\perp{^8}}$. Similar 
counting can be applied to estimate the 
$\kappa_\perp$ dependence for  diffraction of a nucleon into three jets.

\section{Other amplitudes}
So far we have emphasized that the amplitude we computed in 1993 
is calculable using perturbative QCD.
However there are four different contributions which occur at the same order
of $\alpha_s$. The previous term in which the  interaction with the
target gluons follows the gluon-exchange represented  by  the potential $V^g$ 
in the pion wave function has been denoted by  $T_1$. But there is also a  
term, in which 
the interaction with the target gluons occurs  
before the action of $V^g$ is denoted as $T_2$, see Figure~2.
However, the two 
gluons
from the nuclear target can also be annihilated by the exchanged gluon (color 
current of the pion wave function). This
amplitude, denoted as $T_3$, is shown in Figure~3.
The  sum of diagrams
when one target gluon is attached before the potential $V^g$  and  a second
after the  potential $V^g$, see E.g. Figure ~4,
corresponds to an amplitude, $T_4$.

We briefly discuss each of the remaining terms $T_2,T_3,T_4$.
Their detailed evaluation 
will appear in a later publication. 
However we state at the outset that each of these amplitudes 
is suppressed by color coherent effects, and that  each has the same 
$\kappa_\perp^{-4}$ dependence. 
The existence of the term
$T_2$, which uses an interaction that varies as $b^2$,
               caused Jennings \& Miller\cite{jm}
 to worry that the value of
${\cal M}_N$ might be severely reduced due to a nearly complete 
cancellation. Our preliminary 
and incomplete
estimate obtained by neglecting the   term arising from differentiation 
of the potential, $V^g$
finds 
instead enhancement.

 The $T_3$ or  meson-color-flow term
arises from the  attachment of both target gluons to
the gluon appearing in $V^g$ as well as the 
sum of diagrams where one target gluon is
attached to potential $V^g$  in the pion wave function
and another gluon is attached
to a quark. This term is suppressed by color coherent destructive interference
caused by the color neutrality of the $q\bar{q} g$ intermediate state. Thus 
this term has a form which is very similar to that of $T_1$ and $T_2$.
The  $T_4$ term
arises from the sum of diagrams when one target gluon interacts
with a quark in the pion wf before exchange by potential $V^g$
and second gluon interacts after that. The sum of these 
diagrams seems to be
O
because, for our kinematics, the $q$ and 
$\bar q$ 
in the initial and final states 
are not causally connected in these diagrams. The 
mathematical origin of
this near 0  arises from the sum of diagrams having  the form
of contour integral: $\int d\nu \frac{1}{(\nu-a-i\epsilon)(\nu-b-i\epsilon)},$
 which 
vanishes because one can integrate using a closed contour in the lower half 
complex $\nu$-plane.


\section{Summary Discussion}
The purpose of this paper has been to show how to apply  leading-order
perturbative QCD to computing the scattering amplitude for the process:
$\pi
N\to JJ$.  The high momentum component of the  pion wave function, computable 
in perturbation theory 
 is an essential element of the amplitude. 
Another essential feature  of our result (20) 
is the $\sim {1\over \kappa_\perp^4}$ dependence of the amplitude
manifest in a $\kappa_\perp^{-8}$ behavior of the cross section. This feature
needs to be observed experimentally before one can be certain that
the experiment \cite{danny} has verified the prediction of Ref.~\cite{fms93}.
\section{Aknowledgements}
It is a pleasure to dedicate this work to Prof. Kurt Haller, who has been
recently interested in color transparency\cite{kh1}, and  who has long  been
interested in the fundamentals of QCD as applied to light-cone
physics\cite{kh2}. Happy birthday, Kurt,
and our best wishes for many
more to come.

This work has been supported in part by the USDOE.

  \begin{figure}
    {Figure 1.
      Contribution to $T_1$. The high momentum component of the pion interacts
      with the two-gluon field of the target. Only a single diagram of the
      four that contribute is   shown.} 
    \end{figure}
\begin{figure}
    {Figure 2. Contribution to $T_2$. The high momentum component of the final
      $q\bar q$ pair  interacts
      with the two-gluon field of the target. Only a single diagram of the
      four that contribute is  shown.} 
    \end{figure}
\begin{figure}
    {Figure 3. Contribution to $T_3$. The exchanged gluon  interacts
      with the two-gluon field of the target. Only a single diagram of the
      several  that contribute is  shown.} 
    \end{figure}
 \begin{figure}
    {Figure 4. Contribution to $T_4$. A  gluon interacts with a quark
      and another with the  exchanged gluon. Only a single diagram of the
      several  that contribute is  shown.} 
    \end{figure}   
  
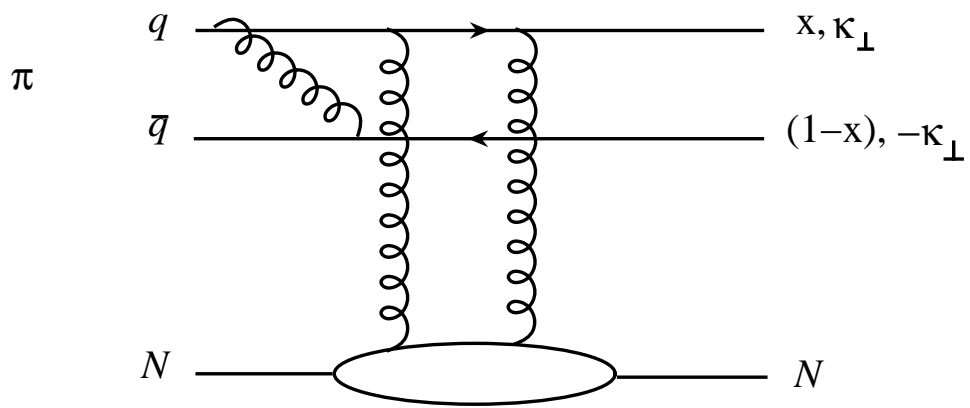
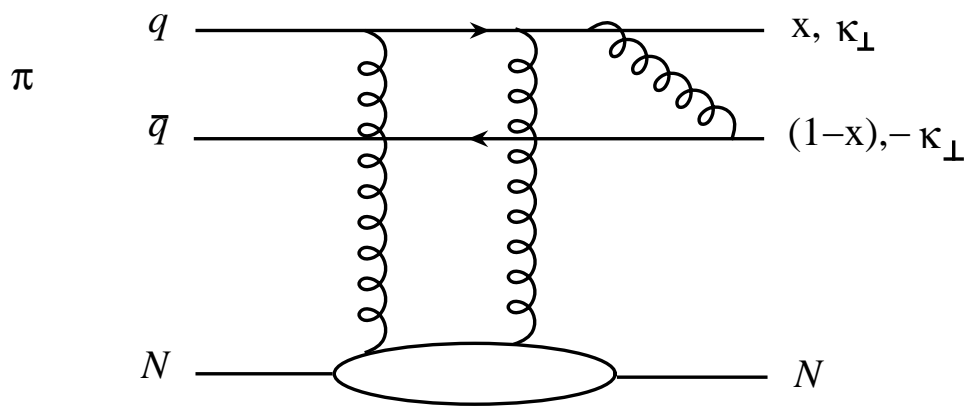
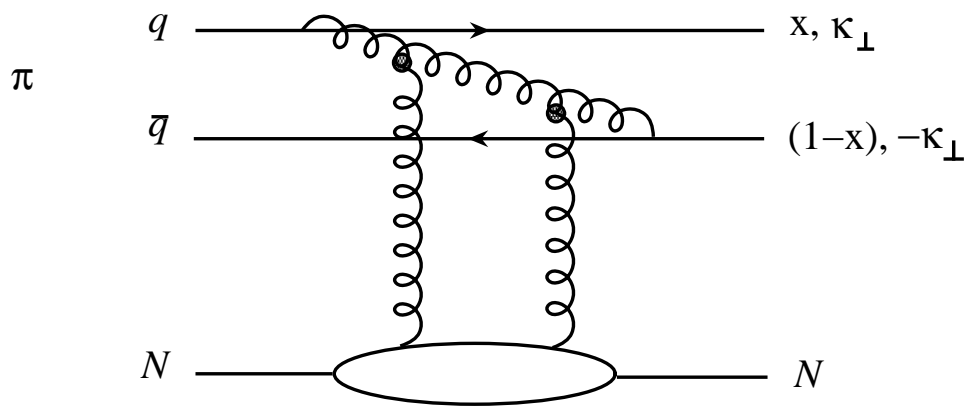
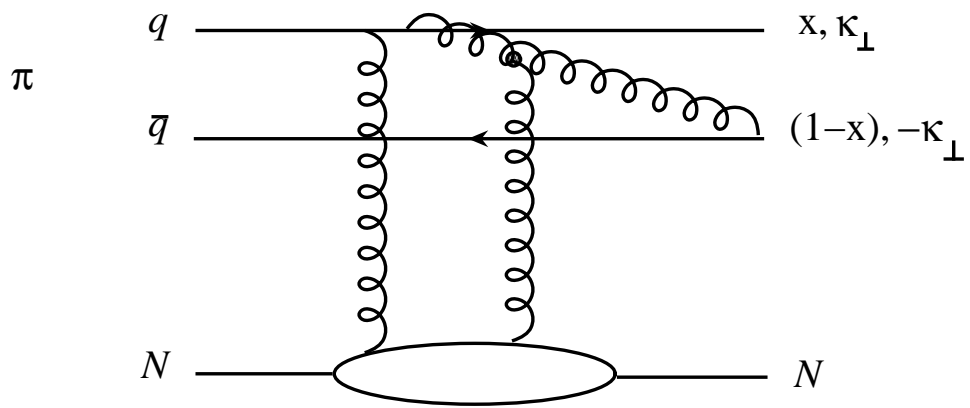
\end{document}